\documentclass[prl,twocolumn,superscriptaddress,showpacs,showkeys]{revtex4}

\usepackage{amsfonts}
\usepackage{amsmath}
\usepackage{amssymb}
\usepackage{graphicx}
\usepackage{dcolumn}
\usepackage{bm}
\usepackage{hyperref}

\begin{document}
\title{Theory of giant magnetoresistance at misfit interfaces}
\author{Daichi Asahi}
\affiliation{Department of Applied Physics, University of Tokyo, Tokyo 113-8656, Japan}
\author{Naoto Nagaosa}
\affiliation{Department of Applied Physics, University of Tokyo, Tokyo 113-8656, Japan}
\affiliation{RIKEN Center for Emergent Matter Science, ASI, RIKEN, Wako 351-0198, Japan}
\date{\today}

\begin{abstract}
We study theoretically the resistance  at the interface between the
two planar systems with different lattice constants $a$ and $b$.
The resistance and the effect of the magnetic field depends sensitively on the ratio $a/b$. The size of the enlarged unit cell $\lambda= n_A a = n_B b$ 
($n_A,n_B $: integers) is the crucial quantity,
and the magnetic flux penetrating this enlarged unit cell determines 
the oscillation of the 
resistance. Therefore, the magnetoresistance is very much
enhanced at (nearly) incommensurate relation between $a$ and $b$.
\end{abstract}
\maketitle

Interfaces between different materials are the sources of rich physics and 
functions. Novel phenomena emerge that are not expected from each of the
constituents~\cite{Hwang}. One example is a two dimensional metallic state appearing 
at the interface between two insulators LaAlO$_3$ and SrTiO$_3$~\cite{Ohtomo}. 
Even superconductivity appears in this two-dimensional system, which is
an issue of recent intensive interests~\cite{Hwang}. Another example is the
tunneling magnetoresistance (TMR)~\cite{TMR}. The resistance across the
interface between the two ferromagnets depends strongly on the relative 
direction of the magnetizations. Therefore, transport properties perpendicular 
to the interface offer many useful functions for applications. 

An essential nature of interfaces between different systems is the
misfit of the lattice constants, which often causes the distortion of the
lattice structure to relax this misfit when it is not so large. In this case, 
the lattice constants slowly change from the interface to the bulk region, 
and correspondingly electrons adiabatically follow this gradual change. 
When the misfit is larger, on the other hand, the system can not
remedy this misfit and an incommensurate situation occurs 
at the interface. 

Incommensurate systems attract recent attention from the viewpoints
of charge/spin density waves~\cite{Bak}, localization of
wavefunctions~\cite{Sokoloff}, and
quasi-crystals~\cite{QX}.  In these systems, incommensurability occurs 
in the bulk states, which is rather exceptional or special cases. 
On the other hand, the incommensurability occurs
very often at interfaces since there is no definite relation 
between the lattice constants of the two systems.

\begin{figure}
\begin{center}
\includegraphics[height=8cm,width=5cm,angle=-90]{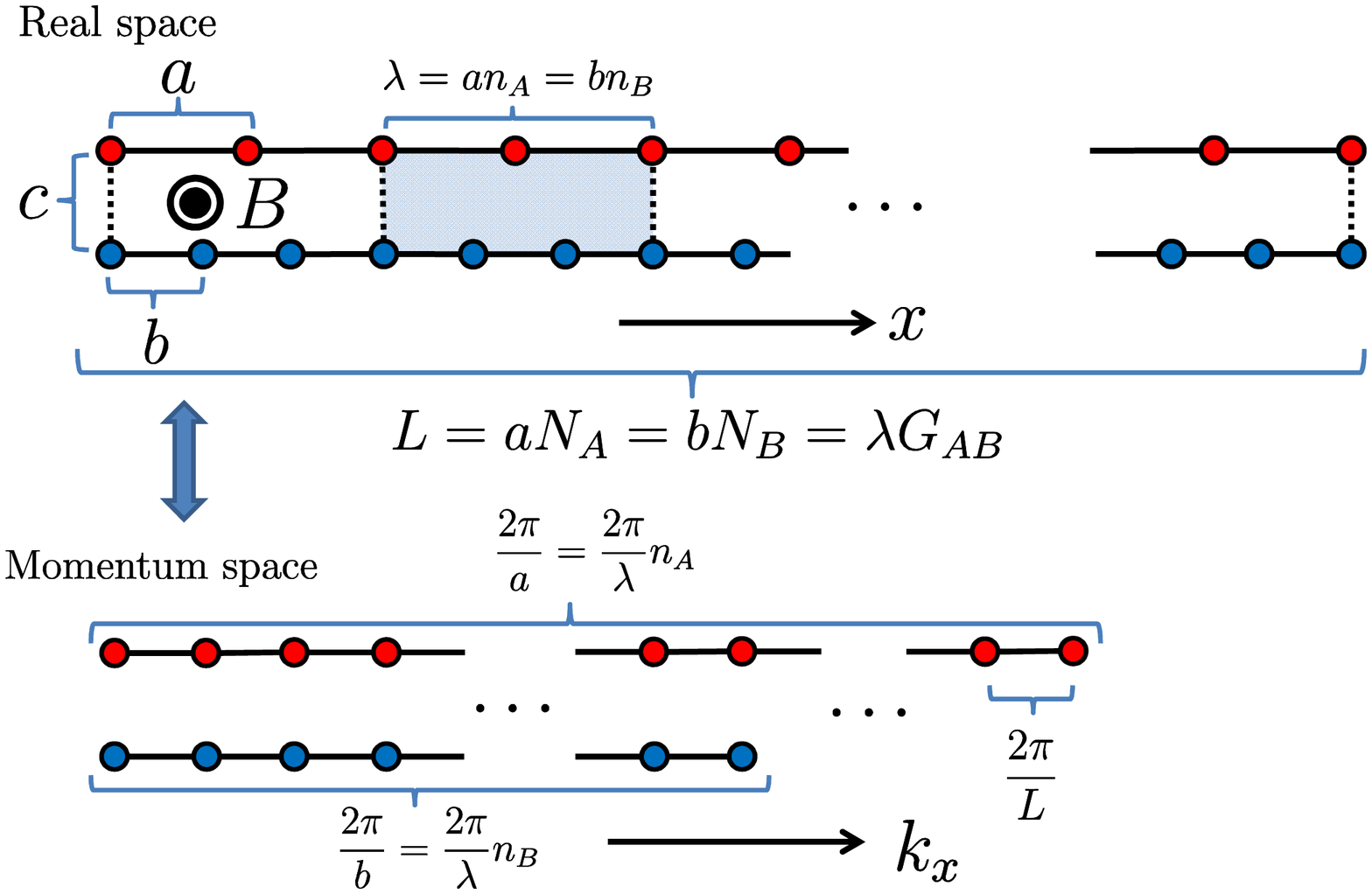}
\caption{
The upper panel indicates two chains in the real space.
One chain has the lattice constant $a$, and the other chain has the 
lattice constant $b$, which are placed in a magnetic field.
The smallest periodic part is constructed from $n_{A}$ 
sites of chain A and $n_{B}$ sites of chain B,
i.e., the size of the enlarged unit cell $\lambda$ is 
given by $\lambda=a n_A = b n_B$.
$G_{AB}$ is the number of this enlarged unit cell and hence is the 
greatest common divisor of $N_{A}$ and $N_{B}$, i.e., 
$N_{A} = n_{A} G_{AB}$ and $ N_{B} = n_{B} G_{AB}$, and $L = G_{AB} \lambda$.
The bottom panel indicates the momentum space of chain $A$ and chain $B$.
The Brillouin zones of $A$ and $B$ are discretized by the same unit $2 \pi/ L$, 
while the sizes are different.
The Brillouin zones are decomposed into $n_{A}$ and $n_{B}$ parts by $\frac{2\pi}{\lambda}$. 
}
\label{fig:lattice}
\end{center}
\end{figure}

In this paper, we study theoretically the tunneling conductance
across the interface between the two two-dimensional systems 
$A$ and $B$ with lattice constants $a$ and $b$, respectively. 
The ratio $a/b$ matters significantly in this tunneling process, and also
its sensitivity to an external magnetic field.
 
Let us start with the two one-dimensional chains $A$ and $B$.
The extension to two-dimensions is straightforward. 
We assume two chains have the same length $L$.
The number of sites in chain $A$ and in chain $B$ are $N_{A}$ and $N_{B}$,
which are determined from $L = a N_A = b N_B$.
We assume that two chains are parallel to $x$-direction 
with the separation $c$, and 
the tunneling amplitude between the $n$th site in chain A 
and the $m$th site in 
chain B is given by 
\begin{equation}
t_{nm}=t_{AB}e^{iBc\frac{an+bm}{2}}
\left(e^{-\frac{\left|an-bm\right|}{d}}
+e^{-\frac{L-\left|an-bm\right|}{d}}\right). 
\label{eq:tunnel}
\end{equation}
Here $d$ characterizes the spatial extent of the tunneling process.
Two chains are placed in a magnetic field, which is perpendicular to the plane 
including two chains.
The magnetic field induces AB phase (Aharanov-Bohm phase).
We choose the gauge as $\bm{A} = (0,Bx)$. 

We rewrite Eq.~(\ref{eq:tunnel}) by wavenumber representation as 
\begin{equation}
t_{kp}  =  \frac{1}{\sqrt{N_{A}N_{B}}}\sum_{nm}t_{nm}
e^{-i\frac{2\pi k}{N_{A}}n+i\frac{2\pi p}{N_{B}}m} \label{eq:tkp},
\end{equation}
where the wavenumbers are specified by the integers $k$ and $p$.
The lattice constant $\lambda$ of the composite system of $A$ and $B$
is given by $\lambda = a n_{A} = b n_{B}$, where we define as 
$n_{A} = \frac{N_{A}}{G_{AB}}$ and $ n_{B} = \frac{N_{B}}{G_{AB}}$. 
$G_{AB}$ is the greatest common divisor of $N_{A}$ and $N_{B}$, and is the
number of the unit cells with the lattice constant $\lambda$, 
i.e., $L = G_{AB} \lambda$. 
The translational symmetry by $\lambda\bm{e}_{x}$ leads to 
the conservation of wave numbers by ${\rm mod}~\frac{2\pi}{\lambda}$.
The Brillouin zones are decomposed into $n_{A}$ and $n_{B}$ 
parts by $\frac{2\pi}{\lambda}$. 

The summations in Eq.~(\ref{eq:tkp}) can be carried out (See in the appendix) 
and $t_{kp}$ is obtained as 
\begin{align}
t_{kp}  & =  \frac{G_{AB}}{\sqrt{N_{A}N_{B}}}\delta_{G_{AB}}\left(k-p-2M\right)f
\left(k-M,p+M\right) \label{eq:tkpL}\\
f\left(k,p\right)& = \frac{\left(1-e^{-\frac{L}{d}}\right)
\sinh\left(\frac{\xi}{d}\right)}{\cosh\left(\frac{\xi}{d}\right)
-\cos\left(\frac{2\pi x_{A}}{N_{A}}k-\frac{2\pi x_{B}}{N_{B}}p\right)}. 
\label{eq:inf} 
\end{align}
$\delta_{G_{AB}}$ is defined as
\begin{equation}
\delta_{G_{AB}}\left(k-p\right)=\begin{cases}
1 & k-p = 0 \mod G_{AB}\\
0 & {\rm otherwise} \label{eq:delta}.
\end{cases}
\end{equation}
Equation~(\ref{eq:delta}) represents the wavenumber conservation.
$\xi$ is the characteristic length of this system, which is represented as 
$\xi=\frac{L}{L_{AB}}=\frac{b}{n_{A}}=\frac{a}{n_{B}}$. 
$x_{A}$ and $x_{B}$ are solutions of the Diophantine equation 
$
n_{B}x_{A}-n_{A}x_{B}=1.
$
This equation has an integer solution because $n_{A}$ and $n_{B}$ are 
coprime.
$(x_{A} + m n_A,x_{B} + m n_{B})$ is also the solution, where $m$ is 
an arbitrary integer, 
but this indefiniteness is not concerned with Eq.~(\ref{eq:tkpL}) 
because of Eq.~(\ref{eq:delta}).
Equation~(\ref{eq:inf}) represents the interference 
in the periodic part.

We consider the resistance at an interface between the two-dimensional 
systems $A$ and $B$, which are the straightforward generalization of the
above chain system. 
The conductance $g$ per one enlarged unit cell is represented as 
\begin{equation}
g =\frac{G}{G_{AB}^2}=\frac{1}{n_{A}^{2}n_{B}^{2}}
\sum_{k_{x},k_{y}=0}^{n_{A}-1}\sum_{p_{x},p_{y}=0}^{n_{B}-1}
\sigma_{\bm{k}\bm{p}}^{n_{A}n_{B}}(\bm{b}). \label{eq:con-sum}
\end{equation}
We take a limit $G_{AB} \rightarrow \infty$, and 
\begin{align}
&\sigma_{\bm{k}\bm{p}}^{n_{A}n_{B}}(\bm{b}) 
= \int_{0}^{1}dx\int_{0}^{1}dy
F\left(\bm{k}+\bm{x}+\frac{\bm{b}}{2},\bm{p}
+\bm{x}+\frac{\bm{b}}{2}\right)
\nonumber \\
&\times \delta
\left(\xi_{A}\left(\frac{2\pi}{n_{A}}\left(\bm{k}
+\bm{b}+\bm{x}\right)\right)\right)
\delta\left(\xi_{B}\left(\frac{2\pi}{n_{B}}
\left(\bm{p}+\bm{x}\right)\right)\right). \label{eq:con-part-con}
\end{align}
with
\begin{align}
F(\bm{k},\bm{p}) & =\left|t_{AB}f_{C}\left(k_{x},p_{x}\right)f_{C}
\left(k_{y},p_{y}\right)\right|^{2} \\
f_{C}\left(k,p\right)&=\frac{\sinh
\left(\frac{\xi}{d}\right)}{\cosh\left(\frac{\xi}{d}\right)
-\cos\left(\frac{2\pi x_{A}}{n_{A}}k-
\frac{2\pi x_{B}}{n_{B}}p\right)} \label{eq:factor}.
\end{align}
$\bm{b}$ is the dimensionless magnetic flux penetrating 
the enlarged unit cell, i.e., 
$\bm{B}c \lambda = 2\pi \bm{b}$.
The resistance $R$ per the unit area 
is given by $R = \frac{\lambda^{2}}{g}$, which is the physical quantity of our 
main interest.    

\begin{figure}
\begin{center}
\includegraphics[height=8cm,width=8cm]{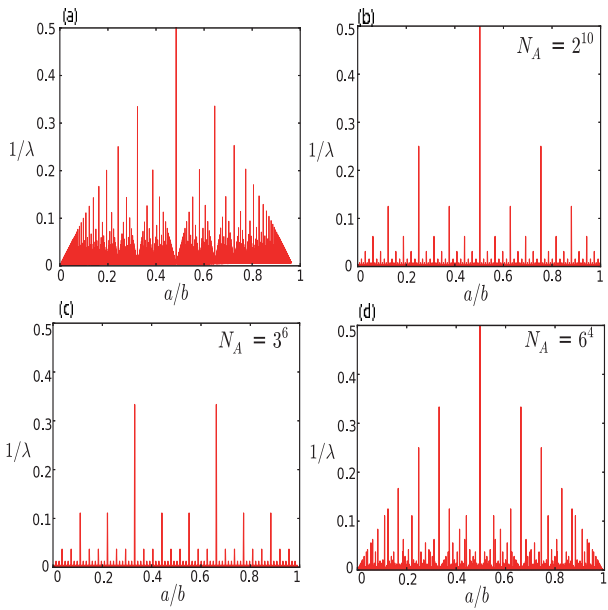}
\caption{The variation of $\frac{1}{\lambda}$ ($\lambda$ the 
size of the enlarged
unit cell) as a function of $\frac{a}{b}=\frac{N_{B}}{N_{A}}$:
Panel (a) indicates the global behavior while
the rest panels (b),(c), and (d) indicate $\frac{1}{\lambda}$ for 
selected $\frac{a}{b} = \frac{N_{B}}{N_{A}}$
with fixed $N_{A} = 2^{10},3^{6}$, and $6^{4}$, respectively.
When $N_{A} = 2^{10},3^{6}$, it show fractal structures, 
but it shows a complex structure when $N_{A} = 6^{4}$.
}
\label{fig:lambda}
\end{center}
\end{figure}

First, we study the lattice constant $\lambda$ of the enlarged unit cell.
$\frac{1}{\lambda}$ change radically as the ratio 
$\frac{a}{b}=\frac{n_B}{n_A}= \frac{N_{B}}{N_{A}}$.
Particularly, $\frac{1}{\lambda}$ indicates a fractal 
architecture with fixed $N_{A}$, if
$N_{A}$ is a power of a prime number.
We show this relation in Fig.~\ref{fig:lambda} with the 
fixed lattice constant $a$ of chain $A$.
The upper left panel indicates the global behavior of 
$\frac{1}{\lambda}$ v.s. $\frac{a}{b}=\frac{N_B}{N_A}$,
and the rest panels show the graphs for selected $\frac{N_{B}}{N_{A}}$
with fixed $N_{A} = 2^{10},3^{8},6^{4}$.
When $N_{A} = 2^{10},3^{8}$, it shows clear fractal structures, while
it shows a complex structure when $N_{A} = 6^{4}$.
When $N_{A} = p^{n}$ where $p$ is a prime number,
$\frac{1}{\lambda}$ is given by the following calculations.
We define
 $S_{i}=\left\{ p^{n-i},2p^{n-i},\cdots
\left(p^{i}-1\right)p^{n-i}\right\}  \enskip (i = 1,2,\cdots n)$
and $P_{i} = S_{i}-S_{i-1} \enskip (i = 1,2,\cdots n)$, where we set 
$S_{0} = \phi$ (empty set).
The relation between $\lambda$ and $\frac{N_{B}}{N_{A}}$ is 
specified as 
\begin{equation}
\frac{1}{\lambda}=\frac{1}{ap^{i}} 
\label{eq:lambda}
\end{equation}
when $N_{B}\in P_{i}$.
This relation and the definition of $P_{i}$ 
generate the fractal structure scaled by $\frac{1}{p}$.
When $N_{A}$ is not a power of a prime number,
this fractal structure is not there, but 
$\frac{1}{\lambda}$ shows a highly singular behavior
as a function of $\frac{a}{b}$ as shown in panels (a) and (d) of 
Fig.~\ref{fig:lambda}.
 
Next, we consider the magnetic field dependence of
$R$ to see the proper scaling for $\bm{B}$.
In a limiting case $d \rightarrow 0$,
the hopping amplitude is finite only between sites whose 
$x$-coordinates and $y$-coordinates are the same.
In this limit, $F =1$ and it is clear that the resistivity 
has the period $\Delta b_{i}=1$,
which is independent of $(n_A,n_B)$.
From $\bm{B}c \lambda = 2\pi \bm{b}$, the resistivity of 
the misfit interface is 
enhanced for larger $\lambda$ (or $n_A, n_B$), i.e., 
(nearly) incommensurate case. 
In this case, $\xi=\frac{a}{n_{B}}$ is small, 
so we cannot assume $\xi \gg d$.
Then there is a crossover from $d \to 0 \ll \xi$ to $d \gg \xi$, 
and the interference in the periodic part $F$
occurs in the latter case where the period of the conductivity becomes 
$ \Delta b_{i}=2 l_{AB}$, where $l_{AB}$ is the least common multiple of 
$n_{A}$ and $n_{B}$.
However, as will be seen for explicit examples, 
the variation of $R$ occurs within the
scale of $\Delta b_i = 1$ even for $d \gg \xi$.

We estimate the behavior of $R$ with $\lambda$ in the limiting cases.
In a limiting case $d \rightarrow 0$, one enlarged unit cell has one pair 
of sites connected by the finite hopping amplitude, and hence 
$R$ grow as $\propto \lambda^2$. 
When $d \cong a \cong b$, which is more relevant to the realistic systems,
there are many finite hopping amplitudes 
between sites in one enlarged unit cell, and 
the number of site in each chain does not becomes important.
The hopping amplitude per area does not change with $\lambda$ and 
the scale of $R$ is nearly constant although the magnetic field dependence is
sensitive to $\lambda$. 
In the extreme case $a \ll b$, on the other hand, 
the situation is similar to the case of $d=0$,
and $R$ is expected to grow with $\propto \lambda^2$.

Now, let us study a concrete example.
We assume the simplest tight binding model
on a square lattice both for $A$ and $B$ as
$\epsilon_{A(B)}\left(\bm{k}\right)=\cos k_{x}+\cos k_{y}$
with the chemical potential at $\mu=0$, and the
magnetic field is along the $x$-axis, i.e., $b_{y} = 0$.
From Eq.~(\ref{eq:con-sum}) and Eq.~(\ref{eq:con-part-con}),
we can obtain the analytic form of the conductivity as given in the
appendix.
In the following calculations, we fix the lattice constant of 
$A$ as $a=1$ and
set $\frac{t_{AB}^2}{2\pi^2} = 1$.

\begin{figure}
\begin{center}
\includegraphics[height=3cm,width=8cm]{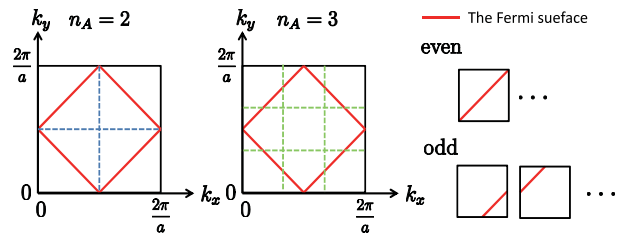}
\caption{The Fermi surface for $\epsilon_{A(B)}\left(\bm{k}\right)=\cos k_{x}+\cos k_{y}$:
The Fermi surfaces are indicated by red lines.
Due to the folding of the 1st Brillouine zone (BZ), the Fermi surface
is segmented into smaller pieces in the reduced BZ, which depends
on the parity of $n_{A}$ ($n_{B}$).
We show two examples $n_A=2$ and $3$ at the right panel
for even and odd $n_A (n_B)$, respectively.
}
\label{fig:model}
\end{center}
\end{figure}
In this model, the Fermi surfaces is straight line as indicated in Fig.~\ref{fig:model}.
At the misfit interface, the Brillouin zone is reduced to 
$n_{A}\times n_{A}$ ($n_{B} \times n_{B}$) 
parts by $\frac{2\pi}{\lambda}$.
The segmented pieces of the Fermi surfaces are determined by 
the parity of $n_{A}$ ($n_{B}$).
This even-odd effect is a special properties of the linear Fermi surfaces.
In Fig.~\ref{fig:model}, we indicate two examples, i.e., cases of $n_A = 2$ and $3$.

We now consider the resistance $R$ 
as a function of the magnetic field $B_{x}$ and the 
dimensionless magnetic flux $b_{x}$.
The resistance $R$ at $B_{x}=b_{x}=0$ reflect 
the even-odd effect of the Fermi surfaces.
When both $n_{A}$ and $n_{B}$ are odd, $R=0$ for
integer values of $b_{x}$.
On the other hand, when either $n_{A}$ or $n_{B}$ is even,
$R=0$ at  $b_{x}= {\rm integer} + \frac{1}{2}$.
Strictly speaking, $R=0$ means the diverging $g$ and hence the
perturbation theory with respect to $t_{AB}$ does not work there.
This divergence has two origins, i.e., the Van Hove singularity and
the fact that the forms of the Fermi surface of $\epsilon_{A}$ 
and $\epsilon_{B}$ are the same.

\begin{figure}
\begin{center}
\includegraphics[height=8cm,width=8cm]{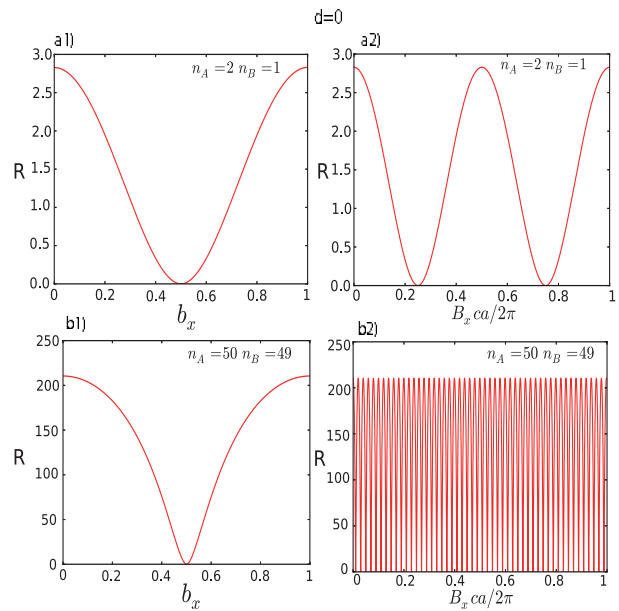}
\caption{
Resistance per unit area $R$ for $d=0$:
The panels (a1) and (b1) shows $R$ as a function of $b_x$
while (a2) and (b2) as a function of $\frac{B_x ac}{2\pi}$.
Panels (a1) and (a2) are for $(n_A,n_B)=(2,1)$, while 
(b1) and (b2) are for $(n_A,n_B)=(50,49)$.
$R$ depends on the magnetic field through the flux penetrating the
enlarged unit cell. and when $\lambda$ is large,
$R$ oscillates as $\frac{B_x ac}{2\pi}$ more rapidly 
and the maximum value becomes larger as $\propto \lambda^2$ as shown in Fig.6(b1).
}
\label{fig:cosP-b}
\end{center}
\end{figure}
At first, we consider a limiting case $d \rightarrow 0$.
The behavior of the resistance $R$ depends on the parity of 
$n_{A}$ and $n_{B}$, 
which reflects the even-odd effect of the Fermi surfaces.
The maximum value of $R$ changes radically in each $n_{A}$ and $n_{B}$, 
which will be discussed later.
$R$ has the period from $b_{x}=0$ to $b_{x}=1$, and
indicates a similar behavior to $b_x$ for all $(n_A, n_B)$ 
except for the even-odd effect.
When either $n_{A}$ or $n_{B}$ is even, the $b_{x}$ is shifted by 
$\frac{1}{2}$. 
$b_{x}$ is the dimensionless magnetic flux penetrating
the enlarged unit cell, which is scaled in an appropriate manner 
in each case as  
$B_x c \lambda = 2\pi b_x$.
The larger $\lambda$ becomes, $R$ oscillates more rapidly as $B_x$. 
The panels (a1) and (b1) of Fig.~\ref{fig:cosP-b} show $R$ as a function of $b_x$,
while (a2) and (b2) shows $R$ as a function of $\frac{B_x ac}{2\pi}$.
Panels (a1) and (a2) are for $(n_A,n_B)=(2,1)$, while (b1) and (b2) 
are for $(n_A,n_B)=(50,49)$.
If $\lambda$ become bigger, the resistivity $R$ oscillate as 
$\frac{B_x ac}{2\pi}$ more rapidly.

\begin{figure}
\begin{center}
\includegraphics[height=8cm,width=8cm]{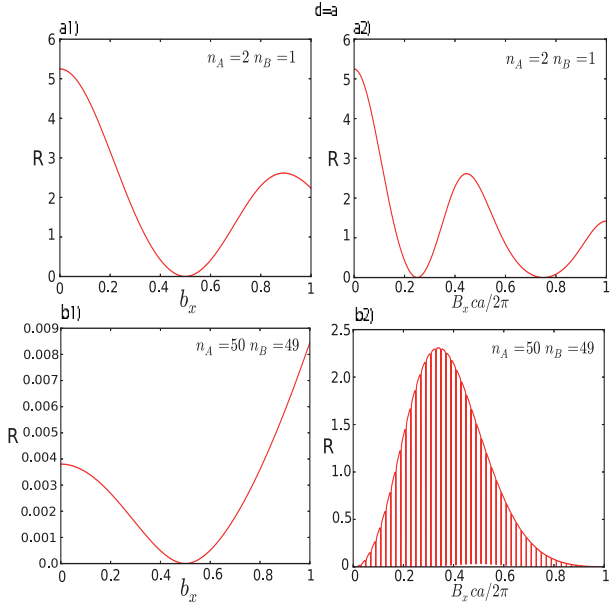}
\caption{
Resistance per unit area $R$ for $d=a$:
The panels (a1) and (b1) shows $R$ as a function of $b_x$
while (a2) and (b2) as a function of $\frac{B_x ac}{2\pi}$.
Panels (a1) and (a2) are for $(n_A,n_B)=(2,1)$, while 
(b1) and (b2) are for $(n_A,n_B)=(50,49)$.
}
\label{fig:cosP-fin}
\end{center}
\end{figure}
Next, we consider a finite $d=a$.
The behavior of the resistance $R$ is not determined only by the parity of 
$n_{A}$ and $n_{B}$,
but it is different in each $n_{A}$ and $n_{B}$ because of the interference term $F$.
The periodicity of $R$ changes from $\Delta b_{x}=1$ to  $\Delta b_{x}=2 l_{AB}$,
but the $b_x$-value at which $R=0$ remains unchanged from the case of $d=0$.
The panels (a1) and (b1) of Fig.~\ref{fig:cosP-fin} show $R$ 
as a function of  $b_{x}$,
while (a2) and (b2) as a function of $\frac{B_x ac}{2\pi}$.
Panels (a1) and (a2) are for  $(n_A,n_B)=(2,1)$, while
(b1) and (b2) are for $(n_A,n_B)=(50,49)$.
In the similar way to the case of $d=0$,
the larger $\lambda$ become, the $R$ oscillates as $b_x$ changes 
of the order of 1 and hence
more rapidly as a function of  $\frac{B_x ac}{2\pi}$.

\begin{figure}
\begin{center}
\includegraphics[height=7cm,width=8cm]{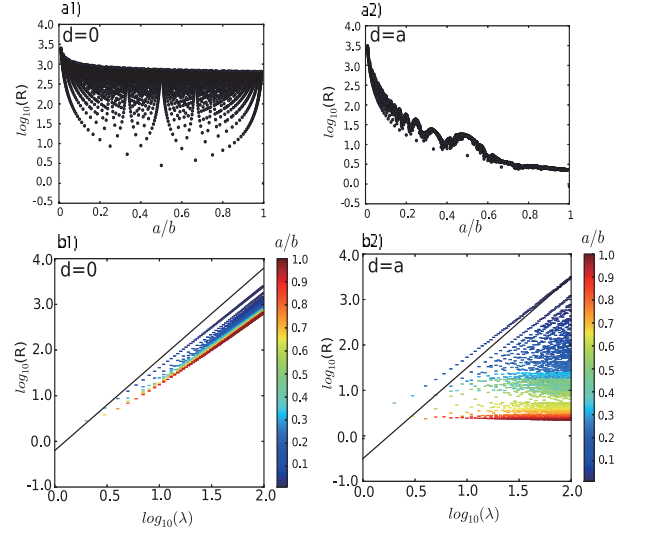}
\caption{
The maximum value of the resistance $R$ 
in the range of $0 \leq B_x ac \leq 2\pi$
as a function of $\frac{a}{b}=\frac{N_B}{N_A}$
for $d=0$ ((a1)) and $d=a$ ((a2)), respectively. 
This complex relations are cleanly organized as a function of $\lambda$,
which are shown in (b1) and (b2).
The black solid lines are the guide to the eyes (slope 2) for the asymptotic relation $R \propto \lambda^2$. 
}
\label{fig:cosP-incom}
\end{center}
\end{figure}

Establishing the enhanced magnetoresistance by $\lambda$,
we next discuss the maximum value of the resistance $R_{\rm max}$
in the region $0 \leq B_x ac \leq 2\pi$.
$R_{\rm max}$ indicates a singular  behavior as a function of
$\frac{a}{b}=\frac{N_{B}}{N_{A}}$ in (a1) and (a2) of Fig.~\ref{fig:cosP-incom} 
for two cases $d=0$ and $d=a$, respectively.
Both show the rather complex and singular behavior, but these 
are neatly organized as a function of $\lambda$ as shown in (b1) and (b2) 
of Fig.~\ref{fig:cosP-incom}. 
When $d \to 0$ (((b1)),  $R_{\rm max}$ increases as $\lambda$ as expected
for all the regions of $a/b$ (asymptotically $\lambda^2$ in the large $\lambda$ limit).
For $d=a$ (((b2)), on the other hand, $R_{\rm max}$ stays almost constant and independent of
$\lambda$ for $a \cong b$, while it approaches to the behavior of $d=0$ as 
$a/b$ decreases.

Now we discuss about the relevance of the present results to real systems. 
The disorder effect at the interface gives the mean free path $\ell$.
When the size $\lambda$ of the enlarged unit cell is larger than
$\ell$, the singular dependence on $a/b$ is broadened. In other words, 
the enhancement of the magnetoresistance saturates by the factor 
$\sim$min$(\ell, \lambda)^2$.
The most relevant case to the real systems is that $a \cong b$ and
$d \cong a$. In this case, the scale of the resistance $R$ per unit area 
does not sensitively depends on the ratio $a/b$, while the 
magnetoresistance is determined by $\lambda$ and depends strongly 
on the ratio $a/b$ in a singular way. The essence of the enhanced 
magnetoresistance is the sensitive change in the interference pattern 
of the wavefunctions within the enlarged unit cell induced by the 
magnetic flux, it is expected that the magnetic field perpendicular 
to the interface also gives the similar effect to the 
parallel case discussed in the present paper.

In summary, we have studied the magentoresistance at the interface with 
misfit of lattice constants. We found that resistance R depends on the 
ratio a/b of the two lattice constants in a singular way, and 
the size $\lambda$ of the enlarged unit cell determines the 
magnitude of the magnetoresistnce, which can he enhanced 
orders of magnitudes when $a/b$ is a (nearly) irrational number.

The authors acknowledge the fruitful discussion with Y. Kawaguchi, K. Burch, M.Kawasaki
and Y.Tokura. 
This work is supported by Grant-in-Aid for Scientific Research
(Grants No. 24224009) 
from the Ministry of Education, Culture,
Sports, Science and Technology of Japan, Strategic
International Cooperative Program (Joint Research Type)
from Japan Science and Technology Agency, and Funding
Program for World-Leading Innovative RD on Science and
Technology (FIRST Program).
\appendix
\begin{widetext}
\section{detail calculations of Eq.~(3)}
In this appendix, we show detail calculations of Eq.~(3).
We write Eq.~(1) by wave-number representation as
\begin{align}
t_{kp} & =  \frac{1}{\sqrt{N_{A}N_{B}}}\sum_{nm}t_{nm}e^{-i\frac{2\pi k}{N_{A}}n+i\frac{2\pi p}{N_{B}}m}\\
 & =  \frac{t_{AB}}{\sqrt{N_{A}N_{B}}}\sum_{nm}e^{-i\frac{2\pi n}{N_{A}}\left(k-M\right)+i\frac{2\pi m}{N_{B}}\left(p+M\right)}\left(e^{-\frac{L}{L_{AB}d}\left|n_{B}n-n_{A}m\right|}+e^{-\frac{L}{d}+\frac{L}{L_{AB}d}\left|n_{B}n-n_{A}m\right|}\right)
\end{align}
,where we use $a=\frac{1}{n_{A}}\frac{L}{G_{AB}}$ and $b=\frac{1}{n_{B}}\frac{L}{G_{AB}}$.
At first, we decompose the summations into $n_A$ and $n_B$ parts by $G_{AB}$,
\begin{align}
t_{kp} & =  \frac{t_{AB}}{\sqrt{N_{A}N_{B}}}\sum_{m_{A}m_{B}=0}^{G_{AB}-1}\sum_{x=0}^{n_{A}-1}\sum_{y=0}^{n_{B}-1}e^{-\frac{L}{L_{AB}d}\left|n_{A}n_{B}(m_{A}-m_{B})+n_{B}x-n_{A}y\right|-i\frac{2\pi\left(n_{A}m_{A}+x\right)}{N_{A}}\left(k-M\right)+i\frac{2\pi\left(n_{B}m_{B}+y\right)}{N_{B}}\left(p+M\right)}\\
 & +  \frac{t_{AB}}{\sqrt{N_{A}N_{B}}}\sum_{m_{A},m_{B}=0}^{G_{AB}-1}\sum_{x=0}^{n_{A}-1}\sum_{y=0}^{n_{B}-1}e^{-\frac{L}{d}+\frac{L}{L_{AB}d}\left|n_{A}n_{B}(m_{A}-m_{B})+n_{B}x-n_{A}y\right|-i\frac{2\pi\left(n_{A}m_{A}+x\right)}{N_{A}}\left(k-M\right)+i\frac{2\pi\left(n_{B}m_{B}+y\right)}{N_{B}}\left(p+M\right)}.
\end{align}
We define $y_{-}={\rm max}\left\{ y:n_{B}x>n_{A}y\right\}$ and 
rearrange the order of the summations, which begins at $y_{-}$, by making use of periodicity.
It is specified as 
\begin{align}
t_{kp} & =  \frac{t_{AB}}{\sqrt{N_{A}N_{B}}}\sum_{m_{A}=0}^{G_{AB}-1}\sum_{x=0}^{n_{A}-1}\sum_{y=0}^{N_{B}-1}e^{-\frac{L}{L_{AB}d}\left(n_{A}(y+y_{-}+1)-n_{B}x\right)-i\frac{2\pi\left(m_{A}n_{A}+x\right)}{N_{A}}\left(k-M\right)+i\frac{2\pi\left(m_{A}n_{B}+y_{-}+1+y\right)}{N_{B}}\left(p+M\right)}\\
 & +  \frac{t_{AB}}{\sqrt{N_{A}N_{B}}}\sum_{m_{A}=0}^{G_{AB}-1}\sum_{x=0}^{n_{A}-1}\sum_{y=0}^{N_{B}-1}e^{-\frac{L}{d}-\frac{L}{L_{AB}d}\left(n_{B}x-n_{A}(y+1+y_{-})\right)-i\frac{2\pi\left(m_{A}n_{A}+x\right)}{N_{A}}\left(k-M\right)+i\frac{2\pi\left(m_{A}n_{B}+y_{-}+1+y\right)}{N_{B}}}.
\end{align}
Here, the summations over $m_{A}$ and $y$ is easily carried out.
$t_{kp}$ becomes
\begin{align}
t_{kp} & = t_{AB} \frac{\delta_{G_{AB}}\left(k-p-2M\right)}{\sqrt{n_{A}n_{B}}}\frac{1-e^{-\frac{L}{d}}}{1-e^{-\frac{Ln_{A}}{L_{AB}d}+i\frac{2\pi}{N_{B}}\left(p+M\right)}}\sum_{x=0}^{n_{A}-1}e^{-\frac{L}{L_{AB}d}\left(n_{A}(y_{-}+1)-n_{B}x\right)-i\frac{2\pi x}{N_{A}}\left(k-M\right)+i\frac{2\pi\left(y_{-}+1\right)}{N_{B}}\left(p+M\right)}\\
 & - t_{AB} \frac{\delta_{G_{AB}}\left(k-p-2M\right)}{\sqrt{n_{A}n_{B}}}\frac{1-e^{-\frac{L}{d}}}{1-e^{\frac{Ln_{A}}{M_{AB}d}+i\frac{2\pi}{N_{B}}\left(p+M\right)}}\sum_{x=0}^{n_{A}-1}e^{-\frac{L}{L_{AB}d}\left(n_{B}x-n_{A}(1+y_{-})\right)-i\frac{2\pi x}{N_{A}}\left(k-M\right)+i\frac{2\pi\left(y_{-}+1\right)}{N_{B}}\left(p+M\right)}.
\end{align}
We define $l$ as $l=\frac{L}{L_{AB}}=\frac{b}{n_{A}}=\frac{a}{n_{B}}$.
$y_{-}$ is equal to the quotient of $n_{B} x$ divided by $n_{A}$.
We denote the remainder as $\Delta x$, i.e., $n_{B} x = n_{A} y_{-} + \Delta_{x}$.
We write $t_{kp}$ by $\Delta x$ and $k-p-2M = n G_{AB}$,
\begin{align}
t_{kp} & = t_{AB} \frac{\delta_{G_{AB}}\left(k-p-2M\right)}{\sqrt{n_{A}n_{B}}}\frac{1-e^{-\frac{L}{d}}}{1-e^{-\frac{b}{d}+i\frac{2\pi}{N_{B}}\left(p+M\right)}}e^{-\frac{b}{d}+i\frac{2\pi}{N_{B}}\left(p+M\right)}\sum_{x=0}^{n_{A}-1}e^{-i\frac{2\pi x}{n_{A}}n+\left(\frac{l}{d}-i\frac{2\pi}{L_{AB}}\left(p+M\right)\right)\Delta x}\\
 & - t_{AB} \frac{\delta_{G_{AB}}\left(k-p-2M\right)}{\sqrt{n_{A}n_{B}}}\frac{1-e^{-\frac{L}{d}}}{1-e^{\frac{b}{d}+i\frac{2\pi}{N_{B}}\left(p+M\right)}}e^{\frac{b}{d}+i\frac{2\pi}{N_{B}}\left(p+M\right)}\sum_{x=0}^{n_{A}-1}e^{-i\frac{2\pi x}{n_{A}}n-\left(\frac{l}{d}+i\frac{2\pi}{L_{AB}}\left(p+M\right)\right)\Delta x}.
\end{align}

We define $x_{A}$ and $x_{B}$ as solutions of the Diophantine equation, which is expressed as
\begin{equation}
n_{B}x_{A}-n_{A}x_{B}=1.
\end{equation}
This equation has a integer solution because $n_{A}$ and $n_{B}$ are coprime.
The general solutions of this equation is represented as 
$(x_{A} + m n_A,x_{B} + m n_{B})$, where $m$ is an arbitrary integer.
Thus $X_{A} = \Delta x x_{A} + m n_A$ and $X_{B} = \Delta x x_{B} + m n_B$ are 
satisfied with
\begin{equation}
n_{B}X_{A}-n_{A}X_{B} = \Delta x.
\end{equation}
There is one-to-one correspondence between $x = 0,1 \cdots n_{A}-1$ and $\Delta x = 0,1 \cdots n_{A}-1$
because $n_A$ and $n_B$ are coprime.
$\Delta x$ is represented as $\Delta x = \Delta x x_{A} + m n_A$ where $m$ is dependent on $x$.
The summation over $x$ is transformed into the summation over $\Delta x$,
\begin{align}
t_{kp} & = t_{AB} \frac{\delta_{G_{AB}}\left(k-p-2M\right)}{\sqrt{n_{A}n_{B}}}\frac{1-e^{-\frac{L}{d}}}{1-e^{-\frac{b}{d}+i\frac{2\pi}{N_{B}}\left(p+M\right)}}e^{-\frac{b}{d}+i\frac{2\pi}{N_{B}}\left(p+M\right)}\sum_{\Delta x=0}^{n_{A}-1}e^{\left(\frac{l}{d}-i\frac{2\pi}{L_{AB}}\left(p+M\right)-i\frac{2\pi n}{n_{A}}x_{A}\right)\Delta x}\\
 & - t_{AB} \frac{\delta_{G_{AB}}\left(k-p-2M\right)}{\sqrt{n_{A}n_{B}}}\frac{1-e^{-\frac{L}{d}}}{1-e^{\frac{b}{d}+i\frac{2\pi}{N_{B}}\left(p+M\right)}}e^{\frac{b}{d}+i\frac{2\pi}{N_{B}}\left(p+M\right)}\sum_{\Delta x=0}^{n_{A}-1}e^{-\left(\frac{l}{d}+i\frac{2\pi}{L_{AB}}\left(p+M\right)+i\frac{2\pi n}{n_{A}}x_{A}\right)\Delta x}.
\end{align}
After the summation over $\Delta x$ and some calculations, $t_{kp}$ become
\begin{equation}
t_{kp} = t_{AB} \frac{\delta_{G_{AB}}\left(k-p-2M\right)}{\sqrt{n_{A}n_{B}}}\left(1-e^{-\frac{L}{d}}\right)\frac{\sinh\left(\frac{l}{d}\right)}{\cosh\left(\frac{l}{d}\right)-\cos\left(\frac{2\pi}{L_{AB}}\left(p+M\right)+\frac{2\pi n}{n_{A}}x_{A}\right)} \label{eq:tkpB}.
\end{equation}
Eq.~(3) is given by arranging Eq.~\ref{eq:tkpB} into the symmetric form by using $k-p-2M = nG_{AB}$,
which is 
\begin{equation}
t_{kp}  =  \frac{t_{AB}}{\sqrt{n_{A}n_{B}}}\delta_{G_{AB}}\left(k-p-2M\right)\left(1-e^{-\frac{L}{d}}\right)\frac{\sinh\left(\frac{l}{d}\right)}{\cosh\left(\frac{l}{d}\right)-\cos\left(\frac{2\pi x_{A}}{N_{A}}\left(k-M\right)-\frac{2\pi x_{B}}{N_{B}}\left(p+M\right)\right)}.
\end{equation}
\section{detail calculations of Eq.~(6) and Eq.~(7)}
In this appendix, we show detail calculations of Eq.~(6) and Eq.~(7).
We assume that the tunneling amplitude between two lattices is the multiple of two copies of Eq.~(1).
The wave-number representation of the tunneling amplitude also become the multiple of two copies of Eq.~(3),
because calculations can be carried out independently in $x$- and $y$-directions.
It is specified as
\begin{align}
&t_{\mathbf{k}\mathbf{p}}=t_{AB}\frac{f_{2D}\left(\bm{k}-\bm{M},\bm{p}+\bm{M}\right)}{n_{A}n_{B}}\delta_{G_{AB}}^{2}\left(\bm{k}-\bm{p}-2\bm{M}\right)  \label{eq:tkp2D} \\
&f_{2D}\left(\bm{k},\bm{p}\right)=f\left(k_{x},p_{x}\right)f\left(k_{y},p_{y}\right) \\
&\delta_{G_{AB}}^{2}\left(\bm{k}-\bm{p}\right)=\delta_{G_{AB}}\left(k_{x}-p_{x}\right)\delta_{G_{AB}}\left(k_{y}-p_{y}\right).
\end{align}
The conductance between two lattices is calculated from 
\begin{equation}
G =  \sum_{\bm{k}\bm{p}}\left|t_{\bm{k}\bm{p}}\right|^{2}\delta\left(\epsilon_{A}\left(\frac{2\pi}{N_{A}}\bm{k}\right)\right)\delta\left(\epsilon_{B}\left(\frac{2\pi}{N_{B}}\bm{p}\right)\right). \label{eq:conductivity}
\end{equation}
$\epsilon_{A(B)}$ is the energy dispersion for lattice A(B), which are periodic by $2\pi$. 
The Brillouin Zones are similarly decomposed into $n_{A} \times n_{A}$ parts and $n_{B} \times n_{B}$ parts by new wave-number conservations by translational symmetries by $\lambda \bm{e}_x$ and $\lambda \bm{e}_y$. 
Eq.~(\ref{eq:conductivity}) is transformed into the more meaningful form 
by the following calculations.
\begin{align*}
G= & \frac{t_{AB}^2}{n_{A}^{2}n_{B}^{2}}\sum_{k_{x},k_{y}=0}^{N_{A}-1}\sum_{p_{x},p_{y}=0}^{N_{B}-1}  \delta_{G_{AB}}^{2}\left(\bm{k}-\bm{p}-2\bm{M}\right)\left|f_{2D}(\bm{k}-\bm{M},\bm{p}+\bm{M})\right|^{2}\delta\left(\xi_{A}\left(\frac{2\pi}{N_{A}}\bm{k}\right)\right)\delta\left(\xi_{B}\left(\frac{2\pi}{N_{B}}\bm{p}\right)\right)\\
= & \frac{t_{AB}^2}{n_{A}^{2}n_{B}^{2}}\sum_{n_{x},n_{y}=0}^{n_{A}-1}\sum_{p_{x},p_{y}=0}^{N_{B}-1}  \left|f_{2D}(\bm{p}+\bm{M}+\bm{n}G_{AB},\bm{p}+\bm{M})\right|^{2}\delta\left(\xi_{A}\left(\frac{2\pi}{N_{A}}\left(\bm{n}G_{AB}+\bm{p}+2\bm{M}\right)\right)\right)\delta\left(\xi_{B}\left(\frac{2\pi}{N_{B}}\bm{p}\right)\right)\\
= & \frac{t_{AB}^2}{n_{A}^{2}n_{B}^{2}}\sum_{n_{x},n_{y}=0}^{n_{A}-1}\sum_{i_{x},i_{y}=0}^{G_{AB}-1}\sum_{m_{x},m_{y}=0}^{n_{B}-1}  \left|f_{2D}(G_{AB}\bm{m}+\bm{i}+\bm{M}+G_{AB}\bm{n},G_{AB}\bm{m}+\bm{i}+\bm{M})\right|^{2}\\
 &   \times\delta\left(\xi_{A}\left(\frac{2\pi}{N_{A}}\left(G_{AB}\bm{n}+G_{AB}\bm{m}+\bm{i}+2\bm{M}\right)\right)\right)\delta\left(\xi_{B}\left(\frac{2\pi}{N_{B}}\left(G_{AB}\bm{m}+\bm{i}\right)\right)\right)\\
= & \frac{t_{AB}^2}{n_{A}^{2}n_{B}^{2}}\sum_{n_{x},n_{y}=0}^{n_{A}-1}\sum_{i_{x},i_{y}=0}^{G_{AB}-1}\sum_{m_{x},m_{y}=0}^{n_{B}-1}  \left|f_{2D}(G_{AB}\bm{n}+\bm{i}+\bm{M},G_{AB}\bm{m}+\bm{i}+\bm{M})\right|^{2}\\
 &   \times\delta\left(\xi_{A}\left(\frac{2\pi}{N_{A}}\left(G_{AB}\bm{n}+\bm{i}+2\bm{M}\right)\right)\right)\delta\left(\xi_{B}\left(\frac{2\pi}{N_{B}}\left(G_{AB}\bm{m}+\bm{i}\right)\right)\right)\\
= & \frac{G_{AB}^{2}}{n_{A}^{2}n_{B}^{2}}\sum_{n_{x},n_{y}=0}^{n_{A}-1}\sum_{m_{x},m_{y}=0}^{n_{B}-1}\sigma_{\bm{n}\bm{m}}^{n_{A}n_{B}}(\bm{M})
\end{align*}
with
\begin{equation}
\sigma_{\bm{k}\bm{p}}^{n_{A}n_{B}}(\bm{M}) = \frac{t_{AB}^2}{G_{AB}^{2}}\sum_{i_{x},i_{y}=0}^{G{}_{AB}-1}\left|f_{2D}(G_{AB}\bm{k}+\bm{i}+\bm{M},G_{AB}\bm{p}+\bm{i}+\bm{M})\right|^{2}\delta\left(\xi_{A}\left(G_{AB}\bm{k}+\bm{i}+2\bm{M}\right)\right)\delta\left(\xi_{B}\left(G_{AB}\bm{p}+\bm{i}\right)\right)
\end{equation}
Here, we take a limit $G_{AB} \rightarrow \infty$.
Each term is replaced by
$\frac{2\pi}{G_{AB}}\bm{i} \rightarrow 2\pi\bm{x}$ and $\frac{4\pi}{G_{AB}}\bm{M} \rightarrow 2\pi\bm{b}$,
\begin{equation*}
\sigma_{\bm{k}\bm{p}}^{n_{A}n_{B}}(\bm{M})
\rightarrow  \iint_{I^2}d^{2}\bm{x}F\left(\bm{k}+\bm{x}+\frac{\bm{b}}{2},\bm{p}+\bm{x}+\frac{\bm{b}}{2}\right)\delta\left(\xi_{A}\left(\frac{2\pi}{n_{A}}\left(\bm{k}+\bm{b}+\bm{x}\right)\right)\right)\delta\left(\xi_{B}\left(\frac{2\pi}{n_{B}}\left(\bm{p}+\bm{x}\right)\right)\right)
\end{equation*}
with
\begin{align*}
F(\bm{k},\bm{p}) & =\left|t_{AB}f_{C}\left(k_{x},p_{x}\right)f_{C}\left(k_{y},p_{y}\right)\right|^{2} \\
f_{C}\left(k,p\right)&=\frac{\sinh\left(\frac{l}{d}\right)}{\cosh\left(\frac{l}{d}\right)-\cos\left(\frac{2\pi x_{A}}{n_{A}}k-\frac{2\pi x_{B}}{n_{B}}p\right)}  \label{eq:factor} \\
I^{2}& = [0,1]\times [0,1].
\end{align*}

\section{detail calculations of the concrete example}
In this section, we calculate the conductivity of the concrete example, i.e.,  
$\epsilon_{A(B)}\left(\bm{k}\right)=\cos k_{x}+\cos k_{y}$ and $b_{y} = 0$.
The conductivity is calculated from
\begin{equation*}
\sigma_{n_{A}n_{B}} =\sum_{k_{x},k_{y}=0}^{n_{A}-1}\sum_{p_{x},p_{y}=0}^{n_{B}-1}\iint_{I^2}d^{2}\bm{x}
\frac{F\left(\bm{k}+\bm{x}+\frac{b_{x}\bm{e_{x}}}{2},\bm{p}+\bm{x}+\frac{b_{x}\bm{e}_{x}}{2}\right)}{n_{A}^{2}n_{B}^{2}}
\delta\left(\epsilon_{A}\left(\frac{2\pi}{n_{A}}\left(\bm{k}+b\bm{e}_{x}+\bm{x}\right)\right)\right)\delta\left(\epsilon_{B}\left(\frac{2\pi}{n_{B}}\left(\bm{p}+\bm{x}\right)\right)\right).
\end{equation*}
The calculation is done by the simple and straightforward method.
We carry on the delta functions by the rule
\begin{equation}
\delta\left(g\left(x\right)\right)=\sum_{i}\frac{\delta\left(x-x_{i}\right)}{\left|g^{'}\left(x_{i}\right)\right|}
\end{equation}
one by one, where $x_{i}$ is the $i$th zero point of $g\left(x\right)$.
The result is 
\begin{align}
\sigma_{n_{A}n_{B}} & =
\frac{1}{4\pi^{2}n_{A}n_{B}}
\sum_{k_{x}=0}^{n_{A}-1}\sum_{p_{x}=0}^{n_{B}-1}
\left(
G\left(2k_{x},2p_{x}\right)+G\left(2k_{x}+1,2p_{x}+1\right)
\right)
\label{eq:con-cosP-fin} \\
G\left(k_{x},p_{x}\right)& =
\frac{F\left(\frac{k_{x}+\Delta}{2},\frac{p_{x}+\Delta}{2},\frac{k_{x}+\Delta-n_{A}+b_{x}}{2},-\frac{p_{x}+\Delta-n_{B}+b_{x}}{2}\right)}{\left|\sin\left(\frac{\pi}{n_{A}}\left(k_{x}+b_{x}+\Delta\right)\right)\sin\left(\frac{\pi}{n_{B}}\left(p_{x}-b_{x}+\Delta\right)\right)\right|}
\end{align}
with $\Delta=\frac{n_{A}+n_{B}}{2}-\left\lfloor\frac{n_{A}+n_{B}}{2}\right\rfloor$.  
In the limit $d \rightarrow 0$,
The conductivity is summarized into the simpler form as 
\begin{equation}
\sigma_{n_{A}n_{B}} =  \frac{1}{2\pi^{2}n_{A}n_{B}}\sum_{k_{x}=0}^{n_{A}-1}\sum_{p_{x}=0}^{n_{B}-1}\frac{1}
{\left|
\sin\left(\frac{\pi}{n_{A}}\left(k_{x}+\Delta+b_{x}\right)\right)
\sin\left(\frac{\pi}{n_{B}}\left(p_{x}+\Delta-b_{x} \right)\right)
\right|}
\end{equation}
\end{widetext}


\begin{thebibliography}{45}%
\bibitem{Hwang}
H.Y. Hwang et al., Nat. Mat. {\bf 11(2)}, 103 (2012).

\bibitem{Ohtomo}
A. Ohtomo,  and H. Y. Hwang, Nature {\bf 427(6973)}, 423 (2004).

\bibitem{TMR}
J.S. Moodera et al., Phys. Rev. Lett. {\bf 74},  3273 (1995).

\bibitem{Bak} P. Bak, Rep. Prog. Phys. {\bf 45}, 587 (1982).
  
\bibitem{Sokoloff} J.B. Sokoloff, Phys. Rep. {\bf 126}, 189 (1985).

\bibitem{QX}E.L. Albuquerque, and M.G. Cottam, 
Phys. Rep. {\bf 376}, 225 (2003).
\end{thebibliography}
\end{document}